\documentclass[12pt,aps,reprint,groupedaddress,showpacs,nofootinbib,prd]{revtex4-1}
\usepackage{amsmath,amssymb,dsfont,mathrsfs,graphicx,slashed}

\usepackage[bottom]{footmisc}

\newcommand{\sbsn}{\slashed{\nabla}}

\usepackage[bottom]{footmisc}
\usepackage{leftidx}
\newcommand{\f}{U}
\newcommand{\bsn}{\boldsymbol{\nabla}}
\newcommand{\rE}{\textrm{Re}({\tilde{\E}})}
\newcommand{\rbE}{\textrm{Re}({\bsn\tilde{\E}})}

\newcommand{\rsbE}{\textrm{Re}({\sbsn\tilde{\E}})}

\newcommand{\ibE}{\textrm{Im}({\bsn\tilde{\E}})}
\newcommand{\ibbE}{\textrm{Im}({\bsn^2\tilde{\E}})}
\newcommand{\isbE}{\textrm{Im}({\sbsn\tilde{\E}})}
\newcommand{\isbbE}{\textrm{Im}({\sbsn^2\tilde{\E}})}
\newcommand{\E}{\mathscr{E}}

\begin{document}

\title{Ernst formulation of axisymmetric fields in $f(R)$ gravity: applications to neutron stars and gravitational waves}

\author{Arthur George Suvorov}
\email{suvorova@student.unimelb.edu.au}
\affiliation{School of Physics, University of Melbourne, Parkville VIC 3010, Australia}
\author{Andrew Melatos}
\email{amelatos@unimelb.edu.au}
\affiliation{School of Physics, University of Melbourne, Parkville VIC 3010, Australia}


\date{\today}

\begin{abstract}
The Ernst formulation of the Einstein equations is generalised to accommodate $f(R)$ theories of gravity. It is shown that, as in general relativity, the axisymmetric $f(R)$ field equations for a vacuum spacetime that is either stationary or cylindrically symmetric reduce to a single, non-linear differential equation for a complex-valued scalar function. As a worked example, we apply the generalised Ernst equations to derive a $f(R)$ generalisation of the Zipoy-Voorhees metric, which may be used to describe the gravitational field outside of an ellipsoidal neutron star. We also apply the theory to investigate the phase speed of large-amplitude gravitational waves in $f(R)$ gravity in the context of soliton-like solutions that display shock-wave behaviour across the causal boundary.
\end{abstract}


\maketitle

\section{Introduction}
Since their initial presentation, the Einstein equations have been rewritten in many different ways \cite{gripod,matzner}. For stationary and axisymmetric spacetimes, the Ernst formulation has proved to be an especially useful representation \cite{ernst}. Ernst showed that it is possible to reduce the Einstein equations in vacuum to a non-linear partial differential equation for a single, complex-valued, scalar function of the spacetime coordinates. The real and imaginary components of a solution to the Ernst equation encode the metric coefficients, which satisfy the Einstein equations by construction. The Ernst formulation offers several advantages \cite{haris}. For example the multipole moments of the spacetime can be read directly off the Ernst variable \cite{fodor,sotapo,pappas}, and new solutions can be generated from old solutions using Kinnersley and other transformations \cite{kinner,sibgat,sibpap}. As well as offering analytic advantages, many numerical techniques are better suited to solving the scalar Ernst equation rather than the tensorial Einstein system \cite{klein}.

Ernst went on to show that this formulation extends to the Einstein-Maxwell theory \cite{ernst2},  where the field equations can be reduced to two equations for two complex-valued scalar functions, one for the metric and one for the electromagnetic 4-potential. Recent work has also shown that general relativity (GR) in higher dimensions can be moulded into a similar `Ernst' form \cite{cosernst}, as can some Brans-Dicke theories \cite{scalartensor,scalartensor2}. Similarly, GR metrics which are cylindrically symmetric but time dependent can be cast into an Ernst form by performing a Wick rotation \cite{guill,korot,wickquev}. It is therefore logical to ask whether or not this formulation extends to other general theories of gravitation. In this paper we show that the formulation extends to $f(R)$ gravity in a natural way; see Ref. \cite{felice} for a review of $f(R)$ theories. It turns out that additional nonlinearities appear in the Ernst equation related to the function $f$ and its derivatives, as well as the Ricci scalar $R$ and its derivatives.

There are two flavours of Ernst equation that we generalise here to the $f(R)$ theory of gravity: stationary, and cylindrically symmetric. Stationary spacetimes arise in numerous physically important contexts; for example, they represent the geometry surrounding a rotating compact object. In particular, the Kerr metric falls into this class, as does the $f(R)$-Kerr-Newman metric \cite{suvmel}, and other deformed-Kerr solutions \cite{johpsa,joh22}. It is important to understand how compact bodies behave in non-GR theories for a variety of reasons, such as testing if GR breaks down in the strong field regime \cite{babak}. Cylindrically symmetric solutions are also valuable; for example, they include cosmological and gravitational wave solutions. In particular, it has been known for a long time that gravitational waves propagate at the speed of light in GR. It has also been shown in linearised $f(R)$ theory that gravitational waves satisfy the Klein-Gordon equation and thus propagate with frequency dependent phase velocities \cite{capmass,berrygair}. However, few results are known regarding the fully non-linear case. In this paper we use the cylindrical $f(R)$ Ernst equation to analyse this problem further. We show that exact, non-linear cylindrical gravitational waves in vacuum $f(R)$ gravity obey non-linear wave equations with dissipative and forcing terms related to the function $f$. From these equations a phase speed can be derived.

As in GR, the $f(R)$ Ernst equations are derived using ``point-like"\footnote{To the authors' knowledge, this terminology was introduced by de Ritis et al. \cite{exactcos,ritis2} and refers to a procedure whereby one associates a Lagrangian with the configuration space spanned by the independent components of $g_{\mu \nu}$ instead of the physical spacetime parametrised by coordinates. In this way one obtains a system depending on only a finite number of degrees of freedom \cite{arnold}.} Lagrangian techniques \cite{exactcos,ritis2,arnold,cap1,sebzer}, which we revisit for the Papapetrou metric and its Wick-rotated counterpart in Section II. In Section III, the Ernst-like equations of motion for the metrics in Section II are derived and are shown to reduce to their GR counterparts when $f(R) = R$. Equipped with the generalised Ernst equations, we work through a simple, formal example in Section IV to demonstrate how one may use the Ernst formulation to derive new exact metrics. This idealised example is potentially useful for studying ellipsoidal compact objects like neutron stars, although its utility is mainly formal at the time of writing. In Section V we use the time-dependent Ernst equation to investigate some properties of large-amplitude gravitational waves in $f(R)$ theories, in particular their speed of propagation. The results are discussed in Section VI.

\section{Equations of motion in $f(R)$ gravity}

We derive the point-like Lagrangian associated with a stationary spacetime in Section II A. The formalism for the cylindrically symmetric case, which is completely analogous, is covered in Section II B.

\subsection{Stationary spacetime}

Following Ernst \cite{ernst,ernst2}, we consider a stationary, axisymmetric spacetime endowed with the Weyl-Lewis-Papapetrou line element in Weyl coordinates $\{t,\rho,\phi,z\}$ \cite{pap,chandbh,konop},
\begin{equation} \label{eq:lineel}
ds^2 = \f^{-1}\left[ e^{2 \gamma} \left( d z^2 + d \rho^2 \right) + B^2 d \phi^2 \right] - \f \left( dt - \omega d \phi \right)^{2} ,
\end{equation}
where $\f$, $\omega$, $B$, and $\gamma$ are functions of $\rho$ and $z$ only. In GR, it was shown by Papapetrou that the vacuum Einstein equations imply $B_{,\rho \rho} + B_{,zz} = 0$ (see also below) \cite{pap}. Hence one can always adopt a set of harmonic coordinates $\{t,\bar{\rho},\phi,\bar{z}\}$ with the properties $dz^2 \mapsto d \bar{z}^2$, $d\rho^2 \mapsto d \bar{\rho}^2$, and $B = \bar{\rho}$ \cite{weyl7}. Therefore, in GR, the function $B$ is redundant and the number of free functions reduces to three without loss of generality. In $f(R)$ gravity this transformation is not always possible because the equations governing the variable $B$ are more complicated, so we must use the more general line element \eqref{eq:lineel} \cite{cosernst,azadi,momeni}. It should be noted that the particular form of the metric \eqref{eq:lineel} holds in vacuum, and a more general form may be required when considering arbitrary matter sources.


The $f(R)$ theory of gravity is a natural generalisation of GR, where the Ricci scalar, $R$, appearing in the Einstein-Hilbert action, is replaced by an arbitrary function of this quantity, $f(R)$. The $f(R)$ action reads
\begin{equation} \label{eq:action}
\mathscr{A} = \int d^{4} x \sqrt{-g} f(R).
\end{equation}
\\
Variation with respect to the contravariant metric components $g^{\mu \nu}$ leads to the vacuum $f(R)$ field equations\footnote{Throughout, Greek symbols range over spacetime indices $0,1,2,3$, while Latin indices are reserved for spatial indices $1,2,3$.} (e.g. \cite{capbeyond})
\begin{equation} \label{eq:fofr1}
0 = f'(R) R_{\mu \nu} -  \frac {f(R)} {2} g_{\mu \nu} + \left[ g_{\mu \nu} \square - \nabla_{\mu} \nabla_{\nu} \right] f'(R),
\end{equation}
\\
where $R_{\mu \nu} = R^{\alpha}_{\mu \alpha \nu}$ is the Ricci tensor, and $\square = \nabla_{\mu} \nabla^{\mu}$ symbolises the d'Alembert operator.

Instead of working with the physical spacetime, one can express the action \eqref{eq:action} directly in terms of the configuration variables $\f, \omega, B$, and $\gamma$, and their first derivatives with respect to the spacetime coordinates. For static, spherically symmetric metrics in $f(R)$ gravity, a set of field equations equivalent to \eqref{eq:fofr1} has been derived by configuration space techniques \cite{cap1,sebzer,newtlim}. In our case, we are considering the metric \eqref{eq:lineel}, and so our configuration variables are $U, \omega, B, \gamma$, and their first derivatives with respect to $\rho$ and $z$.

The Ricci scalar is uniquely determined by the metric coefficients. This information can be self-consistently absorbed into the action \eqref{eq:action} by imposing a constraint equation. To this end, we introduce a Lagrange multiplier $\lambda$ \cite{cap1,sebzer,teyssan},
\begin{equation} \label{eq:action2}
\mathscr{A} = \int d^{4} x \sqrt{-g} \left[ f(R) - \lambda \left( R - \bar{R} \right) \right],
\end{equation}
where $\bar{R}$ is the Ricci scalar expressed explicitly in terms of the configuration variables (as opposed to $R$ which is to be thought of as a function of the spacetime coordinates). Variation of the action \eqref{eq:action2} with respect to the configuration variables then leads to the equations of motion subject to the constraint $R = \bar{R}$. For our case, with respect to \eqref{eq:lineel}, we find
\begin{equation} \label{eq:explicitric}
\begin{aligned}
\sqrt{-g}\bar{R} =&   \frac {1} {\f} \bsn B \cdot \bsn \f + \frac {\f^2} {2B} \bsn \omega \cdot \bsn \omega - \frac {3 B} {2 \f^2} \bsn \f \cdot \bsn \f \\
&- 2 \bsn^2 B - 2 B \bsn^2 \gamma + \frac {B} {\f} \bsn^2 \f,
\end{aligned}
\end{equation}
where the operator $\bsn$ forms a $2$-gradient with respect to the embedded $2$-dimensional metric $d\sigma^2 = dz^2 + d \rho^2$, i.e. we have $\bsn \alpha = (\alpha_{,z},\alpha_{,\rho})$ and $\bsn^2 \alpha = \alpha_{,zz} + \alpha_{,\rho \rho}$ for any scalar function $\alpha(z,\rho)$. In equation \eqref{eq:explicitric}, $\bar{R}$ is a function of the configuration variables $U, \omega, B, \gamma$, and their derivatives. We obtain $\lambda = f'(R)$ by varying the action \eqref{eq:action2} with respect to $R$. Any second order terms (e.g. $\f_{,\rho \rho}$) can be removed from the action \eqref{eq:action2} through integration by parts, and total divergence terms may be removed by invoking Gauss's theorem (see \cite{capbeyond} and Appendix A for details). The Lagrangian, being the integrand of the action \eqref{eq:action2}, reads
\begin{widetext}
\begin{equation} \label{eq:lagrangian}
\begin{aligned}
\mathscr{L} =& \frac {e^{2 \gamma} B} {\f} \left[ f(R) - R f'(R) \right] + \frac {f'(R)} {2 B \f^2} \left[ 4 B \f^2 \bsn B \cdot \bsn \gamma + \f^4 \bsn \omega \cdot \bsn \omega - B^2 \bsn \f \cdot \bsn \f \right] \\
&+ \frac {f''(R)} {\f} \Big\{ 2 \f \left[ \bsn R \cdot \bsn B + B \bsn R \cdot \bsn \gamma \right] - B \bsn \f \cdot \bsn B \Big\},
\end{aligned}
\end{equation}
\end{widetext}
where we have made use of relations (A2)--(A4) derived in Appendix A.  Equation \eqref{eq:lagrangian} reduces to the Lagrangian used by Ernst upto ignorable divergence terms (see above equation (4) in Ref. \cite{ernst}) in the special case $f(R) = R, B = \rho$.

The point-like field equations may now be written down in their entirety by varying the Lagrangian \eqref{eq:lagrangian} with respect to $\f, \gamma, \omega$, and $B$, as well as their derivatives. The equations of motion for $\f$ and $\omega$ \cite{arnold},
\begin{equation} \label{eq:field1}
0 = \frac {\partial \mathscr{L}} {\partial \f} - \frac {\partial} {\partial x^{i}} \frac {\partial \mathscr{L}} {\partial \f_{,i}} ,
\end{equation}
and
\begin{equation} \label{eq:field2}
0 = \frac {\partial \mathscr{L}} {\partial \omega} - \frac {\partial} {\partial x^{i}} \frac {\partial \mathscr{L}} {\partial \omega_{,i}}  ,
\end{equation}
are not written down explicitly here, because they are presented in a simpler form in Section III. Variation of $\mathscr{L}$ with respect to $\gamma$, 
\begin{equation} \label{eq:field4}
0 = \frac {\partial \mathscr{L}} {\partial \gamma} - \frac {\partial} {\partial x^{i}} \frac {\partial \mathscr{L}} {\partial \gamma_{,i}} ,
\end{equation}
yields an integrability condition for the $f(R)$ theory and not a differential equation for $\gamma$, because the Lagrangian \eqref{eq:lagrangian} depends only linearly on derivatives of $\gamma$. If we have $R f'(R) = f(R)$, as in GR, $\gamma$ becomes a cyclic coordinate\footnote{Something similar happens in spherical symmetry; see $\S 3$ of \cite{cap1}.} for the Lagrangian $\mathscr{L}$. Evaluating \eqref{eq:field4} explicitly we find
\begin{widetext}
\begin{equation} \label{eq:field5}
\begin{aligned}
0 = \frac {e^{2 \gamma} B} {\f} \left[ R f'(R) - f(R) \right] + f'(R) \bsn^2 B + f''(R) \left[ 2 \bsn B \cdot \bsn R + B \bsn^2 R \right] + B f'''(R) \bsn R \cdot \bsn R .
\end{aligned}
\end{equation}
\end{widetext}
Equation \eqref{eq:field5} demonstrates a significant difference between theories with $f(R) \neq R$ and GR. When one has $R f'(R) \neq f(R)$, $\gamma$ can be deduced from the variables $\f$ and $B$ by inverting equation \eqref{eq:field5}. As a result, in some ways, the $f(R)$ field equations admit a simpler structure than GR for the metric \eqref{eq:lineel}. In GR, equation \eqref{eq:field5} reads $\bsn^2 B = 0$, and does not constrain $\gamma$. However, since the Ricci scalar must be fixed as zero in GR, equation \eqref{eq:explicitric} fills the role of a differential equation for $\gamma$ given $\f$, $\omega$, and $B$ (solved for through equations \eqref{eq:field1}, \eqref{eq:field2}, and \eqref{eq:field5}, respectively) subject to appropriate boundary conditions. In either case, we have four equations in four variables; see also \cite{ryan} and equations (13.8) in Ref. \cite{gripod}. 
 
After some manipulations, the equation of motion for $B$,
\begin{equation} \label{eq:bfield}
0 = \frac {\partial \mathscr{L}} {\partial B} - \frac {\partial} {\partial x^{i}} \frac {\partial \mathscr{L}} {\partial B_{,i}} ,
\end{equation} 
reads,
\begin{equation} \label{eq:bfield2}
\begin{aligned}
0 =& e^{-2\gamma} B \Big[f'(R) \bsn^2 B - f''(R) \bsn B \cdot \bsn R  \\
&- 2 f''(R) B \bsn^2 R - 2 B f'''(R) \bsn R \cdot \bsn R \Big] +\frac {B^2 f(R)} {\f} .
\end{aligned} 
\end{equation}
\\
\\
In the GR limit, equation \eqref{eq:bfield2} also reduces to $\bsn^2 B = 0$. For GR with nonzero cosmological constant, where we have $f(R) = R - 2 \Lambda$, equation \eqref{eq:bfield2} reads
\begin{equation} \label{eq:somebeq}
e^{-2 \gamma} \f \bsn^2 B + 2 \Lambda B = 0,
\end{equation}
which is a Helmholtz equation for $B$ \cite{santos,astorino}. In this case, equation \eqref{eq:field5} is again identical to \eqref{eq:somebeq}, and the degrees of freedom in the system are reduced self-consistently; $R$ is still fixed (with value $R = 4 \Lambda$), so again \eqref{eq:explicitric} is an equation for $\gamma$ rather than for $R$, and equation \eqref{eq:somebeq} becomes redundant.

Equation \eqref{eq:bfield2} demonstrates the importance of keeping the function $B$ in the line element \eqref{eq:lineel} in general for $f(R)$ gravity. If we were to take $B = \rho$, equations \eqref{eq:field5} and \eqref{eq:bfield2} immediately tell us that there are no $f(R)$ solutions parametrisable by the Papapetrou metric \eqref{eq:lineel} that admit $R = R_{0} = \textrm{constant} \neq 0$ unless $f(R_{0}) = R_{0} f'(R_{0}) = 0$. It is well known that $f(R)$ gravity with $R = R_{0} = \textrm{constant}$ is equivalent to the Einstein equations with effective cosmological constant $\Lambda_{\textrm{eff}} = \frac {f(R_{0})} {2 f'(R_{0})}$ \cite{felice}, provided that $f'(R_{0}) \neq 0$. Therefore, there are no GR solutions with $\Lambda_{\textrm{eff}} \neq 0$ for $B = \rho$. 

In an $f(R)$ theory where $R$ is not constant, equation \eqref{eq:field5} can be used to eliminate $\gamma$ from equation \eqref{eq:bfield2} resulting in an equation relating $B$ and $f$ that reads
\begin{widetext}
\begin{equation} \label{eq:nogammab}
\begin{aligned}
0 =& R f'(R) \left[  \frac  {2 f(R) \bsn^2 B} {R} - f'(R) \bsn^2 B + f''(R) \bsn B \cdot \bsn R  + 2 B f''(R) \bsn^2 R + 2 B f'''(R) \bsn R \cdot \bsn R  \right] \\
&- B f(R) \left[ f''(R) \bsn^2 R - f''(R) \frac {\bsn B \cdot \bsn R} {B} + B f'''(R) \bsn R \cdot \bsn R \right] .
\end{aligned}
\end{equation}
\end{widetext}

Expressions \eqref{eq:field1}, \eqref{eq:field2}, \eqref{eq:field5}, and \eqref{eq:nogammab} are not much simpler than equation \eqref{eq:fofr1}. In Section III we show how one can reduce the expressions obtained above down to a simpler Ernst form.

\subsection{Cylindrically symmetric spacetime}
Consider the Jordan-Ehlers-Kompaneets line element \cite{jordan},
\begin{equation} \label{eq:timelineel}
ds^2 = \f^{-1}\left[ e^{2 \gamma} \left( -d t^2 + d \rho^2 \right) + B^2 d \phi^2 \right] + \f \left( dz - \omega d \phi \right)^{2} ,
\end{equation}
where now $f$, $\gamma$, $B$, and $\omega$ are functions of $t$ and $\rho$. Though we keep the same set of configuration variables, the line element \eqref{eq:timelineel} is of a fundamentally different structure to the Papapetrou metric \eqref{eq:lineel}, and describes different physical scenarios (see Sections IV and V). Following the procedure in the previous section, we find that the integrand of the action \eqref{eq:action2} for the metric \eqref{eq:timelineel} reads
\begin{widetext}
\begin{equation} \label{eq:timelagrangian}
\begin{aligned}
\mathcal{L} =& \frac {e^{2 \gamma} B} {\f} \left[ f(R) - R f'(R) \right] + \frac {f'(R)} {2 B \f^2} \left[ 4 B \f^2 \sbsn B \cdot \sbsn \gamma - \f^4 \sbsn \omega \cdot \sbsn \omega - B^2 \sbsn \f \cdot \sbsn \f \right] \\
&+ \frac {f''(R)} {\f} \left[ 2 \f \left( \sbsn R \cdot \sbsn B + B \sbsn R \cdot \sbsn \gamma \right) - B \sbsn \f \cdot \sbsn B \right],
\end{aligned}
\end{equation}
\end{widetext}
where the complex 2-operator $\sbsn$ acts on scalar functions $\alpha(t,\rho)$ as $\sbsn \alpha = \left(  i \alpha_{,t}, \alpha_{,\rho} \right)$ and $\sbsn^2 \alpha = \alpha_{,\rho \rho} - \alpha_{,tt}$. In particular, the operator $\sbsn$ is formally related to $\bsn$ through the Wick rotation $z \mapsto i t$ (see Ref. \cite{visser} for a discussion of Wick rotations in curved spacetime). Though the line elements \eqref{eq:timelineel} and \eqref{eq:lineel} are different, and are introduced in unconnected contexts, we see that the Lagrangians \eqref{eq:timelagrangian} and \eqref{eq:lagrangian} are equivalent under the Wick rotations $z \mapsto it$ and $t \mapsto -iz$ and the identification $\omega \rightarrow i \omega$.  As a result, the equations of motion, namely equations \eqref{eq:field1}, \eqref{eq:field2}, \eqref{eq:field5}, and \eqref{eq:nogammab}, are also identical to the equations of motion for the metric \eqref{eq:timelineel}, provided one replaces $\bsn$ with $\sbsn$ in each of the expressions and writes $i \omega$ in place of $\omega$ \cite{arnold,guill,korot}. Furthermore, since the operator $\sbsn$ only appears quadratically in the Lagrangian \eqref{eq:timelagrangian}, we have that $\mathcal{L}$ is strictly \textit{real}. Although the operators $\bsn$ and $\sbsn$ are formally related by a complex Wick rotation, the functions $U, \omega, \gamma$, and $B$ appearing in \eqref{eq:timelineel} are real functions of real coordinates. Similar Wick rotation techniques have been applied in the GR case to transform stationary and axisymmetric solutions into cylindrically symmetric and time dependent ones \cite{beck,sanchez,carr}.



\section{Ernst Equation}

\subsection{Stationary spacetime}

The Ernst equation (equation (2) in Ref. \cite{ernst}) is remarkably simple, because the GR Lagrangian \eqref{eq:explicitric} does not depend on the generalised position $\omega$ explicitly, meaning that $\omega$ is a cyclic coordinate, which implies that the associated momentum is conserved \cite{arnold}. This is also true for the $f(R)$ case, as can be seen from expression \eqref{eq:lagrangian}. In particular, the field equation \eqref{eq:field2} reads
\begin{equation} \label{eq:imernst}
0 = \bsn \cdot \left[ \frac {\f^2} {B} f'(R) \bsn \omega \right ] .
\end{equation}
In GR, when $B$ is fixed as $B = \rho$, equation \eqref{eq:imernst} contains the coordinate factor $\rho^{-1}$. Ernst showed that one may introduce a potential function $\varphi$ related to $\omega$ which removes this coordinate dependency \cite{ernst,ernst2}. Such a construction is possible in $f(R)$ theories when $B = \rho$ and is discussed in Appendix B. However, in general, for $B \neq \rho$,  equation \eqref{eq:imernst} is already coordinate independent, because $\bsn$ is defined as the covariant derivative with respect to the 2-metric $d \sigma^2 = dz^2 + d \rho^2$ (and not with respect to the cylindrical 3-metric $d\Sigma^2 = dz^2 + d \rho^2 + B^2 d \phi^2$, which is not flat when $B \neq \rho$, as it is in Ernst's work \cite{ernst,ernst2}). Any coordinate transformations involving $z$ or $\rho$ self-consistently modify the $\bsn$ operator through the Christoffel symbols. As a result, in $f(R)$ gravity, we do not need to, in general, introduce the variable $\varphi$. The reader who is more familiar with the usual GR construction of the Ernst equation involving $\varphi$ can make use of the equations presented in Appendix B to express equation \eqref{eq:imernst} and others in terms of $\varphi$ rather than $\omega$ [see equations \eqref{eq:varphieq} and \eqref{eq:imernst2}]. We elect instead to express our results in terms of $\omega$ to avoid coordinate terms appearing in the general case $B \neq \rho$.


The Ernst equation in GR is obtained by constructing a complex equation, where the vanishing of the real component implies \eqref{eq:field1} and the vanishing of the imaginary component implies \eqref{eq:imernst} \cite{ernst}.


We can obtain an Ernst-type equation for $f(R)$ gravity by introducing a complex-valued function\footnote{Note that the function $\tilde{\E}$ will not be complex differentiable in general since it does not satisfy the Cauchy-Riemann equations (e.g. for static solutions one finds $\omega = 0$ but $\f \neq \text{constant}$ except for the Minkowski spacetime). Both $\f$ and $\omega$ are real and smooth outside of a source in any physically reasonable spacetime, but one must be cautious when seeking to apply complex analysis techniques (e.g. residue theorem) to $\tilde{\E}$.} $\tilde{\mathscr{E}} = \f + i \omega$, making use of equation \eqref{eq:field5}, and recasting both equations \eqref{eq:imernst} and \eqref{eq:field1} into a single equation for $\tilde{\mathscr{E}}$,
\begin{widetext}
\begin{equation} \label{eq:ernst}
\begin{aligned}
0 =&  f''(R) \rE \left[B \rbE - \rE \bsn B \right] \cdot \bsn R + i \left\{ \bsn \left[ \frac {\rE^2} {B} f'(R) \right] \cdot \ibE + \frac {\rE^2} {B} f'(R) \ibbE \right\}  \\
& + f'(R) \left\{ \frac {\rE^4} {B} \ibE \cdot \ibE + \rE \bsn \cdot \left[ B \rbE \right] - \rE^2 \bsn^2 B - B \rbE \cdot \rbE \right\} ,
\end{aligned}
\end{equation}
\end{widetext}
which is to be solved for $\tilde{\E}$ given $B$ and $f$. Equation \eqref{eq:ernst} generalises the Ernst equation to $f(R)$ gravity.

As a consistency check, if we set $f(R) = R$, $B = \rho$, and introduce the potential $\varphi$ through equation \eqref{eq:varphieq} (see Appendix B), then equation \eqref{eq:ernst} reduces correctly to the GR Ernst equation (in our notation)
\begin{equation} \label{eq:grernst}
0 =  \textrm{Re}(\mathscr{E})  \left[\bsn^2 + \frac {\partial_{\rho}} {\rho} \right] \mathscr{E} - \bsn \E \cdot \bsn \E,
\end{equation}
with $\mathscr{E} = \f + i \varphi$. 

To solve the $f(R)$ field equations in practice we may proceed as follows. First, choose an ansatz for the function $f$ and scalar curvature $R$ to investigate the properties of a particular theory of gravity. In principle, equation \eqref{eq:explicitric} can be applied to eliminate $R$ in the Ernst equation \eqref{eq:ernst} and all other equations appearing in Section II. However, if one wishes to look for solutions that are asymptotically flat, specifying a suitably decaying $R$ \textit{a priori} results in a simpler, decoupled system. The linear equation \eqref{eq:nogammab} can be integrated (in principle) to uniquely determine $B$ given any choices of $R$ and $f$. In turn, if $B$ is known, the Ernst equation \eqref{eq:ernst} can be solved for $\f$ and $\omega$. Finally, the remaining metric coefficient $\gamma$ can be immediately determined using equation \eqref{eq:field5}.  The metric is now completely constructed, and one need only check that the constraint equation \eqref{eq:explicitric} holds. If equation \eqref{eq:explicitric} does not hold, the implication is that no $f(R)$ spacetime, parametrisable by the Papapetrou metric \eqref{eq:lineel}, exists for the initial ansatz.

It is worth emphasising that there is a well-studied equivalence between certain $f(R)$ and scalar-tensor theories of gravity \cite{felice,sotscalar} (see \cite{capnoneq} for a dissenting view however). The Ernst equation \eqref{eq:ernst} reduces to known scalar-tensor forms under an appropriate conformal transformation \cite{scalartensor,scalartensor2}. In particular, we recover equations (3.3a)--(3.3c) of reference \cite{scalartensor} and equations (16a) of reference \cite{scalartensor2} (with the exception of the Maxwell fields; see the discussion in Sec. VI) as a subcase of equation \eqref{eq:ernst}, where the $f(R)$ theory is identified with a scalar-tensor theory with a massless scalar field in the Jordan frame (see also Sec. 10.1 of \cite{felice}). We recover the scalar-tensor quadrature relations for $\gamma$,  equations (3.3d)--(3.3e) of \cite{scalartensor}, from equation \eqref{eq:explicitric} together with \eqref{eq:ernst}.

\subsection{Cylindrically symmetric spacetime}
The Ernst formulation derived in the previous section can also be applied to cylindrically symmetric, time-dependent spacetimes. The field equation for $\omega$ under the line element \eqref{eq:timelineel}, which is equivalent to \eqref{eq:field2} under the maps $z \mapsto it$ and $t \mapsto -iz$, reads
\\
\\
\begin{equation} \label{eq:timefield3}
0 = \sbsn \cdot \left[ \frac {\f^2} {B} f'(R) \sbsn \omega \right ] .
\end{equation}
\\
Furthermore, the field equation for $\f$ reads the same as the real part of \eqref{eq:ernst} but with $\sbsn$ in place of $\bsn$ and a sign flip in $\omega$ terms. The Ernst equation for a cylindrically symmetric spacetime in $f(R)$ gravity is then
\begin{widetext}
\begin{equation} \label{eq:timeernst}
\begin{aligned}
0 =&  f''(R) \rE \left[B \rsbE - \rE \sbsn B \right] \cdot \sbsn R + i \left\{ \sbsn \left[ \frac {\rE^2} {B} f'(R) \right] \cdot \isbE + \frac {\rE^2} {B} f'(R) \isbbE \right\}  \\
& + f'(R) \left\{ \frac {-\rE^4} {B} \isbE \cdot \isbE + \rE \sbsn \cdot \left[ B \rsbE \right] - \rE^2 \sbsn^2 B - B \rsbE \cdot \rsbE \right\} ,
\end{aligned}
\end{equation}
\end{widetext}
with $\tilde{\mathscr{E}} = \f + i \omega$. If we let $f(R) = R$, $B = \rho$, and introduce the Wick-rotated potential $\hat{\varphi}$ through \eqref{eq:timephi}, we obtain the equation,
\begin{equation} \label{eq:grernsttime}
0 =  \textrm{Re}(\mathscr{E})  \left[\sbsn^2 + \frac {\partial_{\rho}} {\rho} \right] \mathscr{E} - \sbsn \E \cdot \sbsn \E ,
\end{equation}
which is a known cylindrical variant of the Ernst equation (see equation (22.5) in Ref. \cite{gripod}).

There is an important distinction between equations \eqref{eq:ernst} and \eqref{eq:timeernst}. Since $t$ is a time-like coordinate, the latter equation is hyperbolic, while the former is elliptic. This may have some implications regarding the stability of numerical codes designed to solve such equations (e.g. \cite{celest}). Nevertheless as in Section III A, the real functions $U$ and $\omega$ defining the metric \eqref{eq:timelineel} may be determined from the real and imaginary components of $\tilde{\E}$, respectively.

\section{Worked example: ellipsoidal neutron stars}
We consider here a simple example of an $f(R)$ theory to demonstrate the method presented. Specifically, we search for a solution which generalises the Zipoy-Voorhees metric of GR \cite{zipoy}. The latter metric represents the spacetime exterior to a static compact object which is not spherically symmetric. It tends to the Schwarzschild solution, when the `oblateness' parameter tends to zero. For example, the metric could describe the gravitational field outside a neutron star that, through magnetic or other internal stresses, has become deformed \cite{ruffini,herrera,toktar}.

We begin by assuming that the Ricci scalar takes the simple form
\begin{equation} \label{eq:somericeq}
R = R_{0} \left( \rho^2 + z^2 \right)^{\Gamma},
\end{equation}
where $R_{0}$ is a constant, and we demand either $\Gamma \leq -1$ or $R_{0} = 0$ to obtain an asymptotically flat spacetime\footnote{See the discussion surrounding equation (19) in \cite{newtlim} for a general discussion on sufficient decay conditions required on $R$ for asymptotic flatness.}. The static Zipoy-Voorhees line element takes the form of \eqref{eq:lineel} with the definitions
\begin{equation} \label{eq:zvu}
\f_{\textrm{ZV}} = \left[ \frac {R_{+} + R_{-} - 2 \left(1 + \epsilon\right)/M } {R_{+} + R_{-} + 2 \left(1 + \epsilon\right)/M} \right]^{1+ \epsilon},
\end{equation}
\begin{equation} \label{eq:zvgamma}
\gamma_{\text{ZV}} = \frac {\left( 1 + \epsilon \right)^{2}} {2} \log  \left[ \frac { \left(R_{+} + R_{-} \right)^{2} - 4 \left(1 + \epsilon\right)^{2}/M^2} {4 R_{+} R_{-} } \right],
\end{equation}
\begin{equation} \label{eq:bomega}
\omega = 0,
\end{equation}
\begin{equation}
 B = \rho,
\end{equation}
with
\begin{equation}
R_{\pm} =  \sqrt{ \rho^2 + \left[ z \pm \left(1 + \epsilon\right)/M \right]^2},
\end{equation}
where $M$ is the mass of the object, and $\epsilon$ is the (formally arbitrary) ellipticity parameter. In particular, $\epsilon > 0$ corresponds to an object more oblate than a Schwarzschild black hole, $\epsilon < 0$ corresponds to a more prolate object, $\epsilon = 0$ reduces the metric functions to the Schwarzschild ones, and $\epsilon = -1$ reduces the metric functions to the Minkowski ones \cite{zipoy,herrera}. For the Zipoy-Voorhees metric we have $R_{\mu \nu} = 0$.

One possible way to search for a suitable generalisation of any GR metric is to fix one of the metric functions to be the same as their GR counterpart and see if the structure of the $f(R)$ theory allows for variation in the other metric components. As a simple example, we make the simplifying assumption that $\gamma$ is unchanged from its GR counterpart in \eqref{eq:zvgamma}, i.e. $\gamma = \gamma_{\text{ZV}}$. Searching for solutions where the function $f$ has power-law form \cite{clifbar}
\begin{equation} \label{eq:powerlaw}
f(R) = f_{0} R^{\alpha},
\end{equation}
for some constant $\alpha$, we find that the only possible solutions compatible with equations \eqref{eq:field5} and \eqref{eq:bfield2} are ones with $R_{0} = 0, \bsn^2 B = 0$ and $\alpha \geq 1$.  This result is one of non-existence; for $R_{0} \neq 0$, there does not exist a $\Gamma$ which allows a power-law $f(R)$ solution with $\gamma$ = $\gamma_{\text{ZV}}$ (though there are non-trivial solutions with $R_{0} = 0$ which we derive below). Since $\bsn^2 B = 0$ we may take $B = \rho$, as in GR, without loss of generality.

Suppose we introduce the ansatz
\begin{equation} \label{eq:generalq}
\f = e^{2Q} \f_{\textrm{ZV}},
\end{equation}
for some function $Q$ which tends to zero at infinity (so that $g_{tt}$ tends to unity). The Ernst equation \eqref{eq:ernst} may be written down in full, though the expressions are lengthy, so we avoid them here. However, if we further assume $\alpha > 1$, so that we work within the realm of strictly non-GR theories, then the Ernst equation \eqref{eq:ernst} is satisfied for any choices of $\omega$ and $Q$. As such, we have that the Ernst equation \eqref{eq:ernst}, equation \eqref{eq:nogammab} for $B$, and equation \eqref{eq:field5} for $\gamma$ are all satisfied for the above choices. The remaining equation is the consistency relation for the Ricci scalar, equation \eqref{eq:explicitric}, which forms an eikonal equation for $\omega$,
\begin{equation} \label{eq:eikonal}
\begin{aligned}
\bsn \omega \cdot \bsn \omega =& \frac{ 4 e^{-4Q} \rho^2} {\f_{\textrm{ZV}}^3} \Bigg[ \bsn \f_{\textrm{ZV}} \cdot \bsn Q \\
&+ \f_{\textrm{ZV}} \left( \bsn Q \cdot \bsn Q - \bsn^2 Q - \frac {Q_{,\rho}} {\rho} \right) \Bigg] .
\end{aligned}
\end{equation}
Equation \eqref{eq:eikonal} is subject to Dirichlet boundary conditions, i.e. $\omega$ must vanish at infinity. Clearly $\omega = Q = 0$ is a solution to \eqref{eq:eikonal}, which simply reproduces the Zipoy-Voorhees solution. Equation \eqref{eq:eikonal} suggests that there is a great deal of freedom in obtaining rotating (or static) generalisations of the Zipoy-Voorhees metric in $f(R)$ gravity. It is well known that the Dirichlet eikonal equation \eqref{eq:eikonal} admits unique solutions for $\omega$ for any well-behaved choice of $Q$ (e.g. \cite{eikonal}). As such, there are infinitely many generalisations of the Zipoy-Voorhees metric, each uniquely corresponding to a particular choice of the function $Q$ (in contrast to GR, where the Ernst equation \eqref{eq:grernst} further restricts the choices of $Q$). As an example, if we take 
\begin{equation} \label{eq:someq1}
Q = -\ln \left(1 - \sigma \f_{\textrm{ZV}} \right),
\end{equation}
where $\sigma$ is an arbitrary constant, we obtain another static solution with $\omega = 0$ since the right hand side of \eqref{eq:eikonal} vanishes. As can be verified directly by substitution, the metric given by \eqref{eq:zvu}--\eqref{eq:someq1} does solve the $f(R)$ field equations \eqref{eq:fofr1} with $R = 0$ for any constant $\sigma$, but has non-vanishing Ricci tensor unless $\sigma = 0$.  In the zero ellipticity limit, $\epsilon \rightarrow 0$, we obtain the Reissner-Nordstr{\"o}m metric \cite{lukas}. A physical interpretation of $\sigma$ is not readily available without performing some additional analysis, i.e. by constructing the multipole moments and matching them with a suitable Newtonian solution \cite{suvmel,pappas}. Such an analysis will be performed elsewhere. It is easy to see that the function $U_{\text{ZV}}$ from \eqref{eq:zvu} is bounded for any $\epsilon \geq -1$, and so we may take $\sigma$ small if necessary to ensure that $U_{\text{ZV}} < \sigma^{-1}$ everywhere, so that the presence of $Q$ does not introduce singularities into the spacetime.

It is likely that more general metrics that include the Zipoy-Voorhees metric as limiting cases exist, where the form of the Ricci scalar differs from \eqref{eq:somericeq}. In particular, the choice made in \eqref{eq:somericeq} resulted in the somewhat trivial property $R_{0} = 0$. Several other choices, such as taking the simple exponential $R \propto e^{-(\rho^2 + z^2)}$, appear to lead to the same non-existence result. In any event, the metric given by \eqref{eq:zvu}--\eqref{eq:someq1} can be used to describe the metric exterior to deformed neutron stars in $f(R)$ gravity. The presence of $\sigma$ (and $Q$) indicates that neutron stars are arbitrarily `hairy' in $f(R)$ gravity; parameters other than their mass and angular velocity influence their properties as seen by observers at infinity \cite{suvmel,doneva}. Neutron stars are also known to be hairy in scalar-tensor theories of gravity, so the equivalence between certain Brans-Dicke and $f(R)$ theories supports the conclusions outlined above \cite{papgeo}.


\section{Phase speed of nonlinear Gravitational waves}


In this section we demonstrate a physical application of the Ernst equation \eqref{eq:timeernst} to gravitational waves. In particular, we examine the behaviour of freely propagating, nonlinear gravitational waves (solitons) in a vacuum $f(R)$ theory. Gravitational waves are often studied within the framework of perturbation theory, whereby the linearised theory, valid far away from the source, provides both an equation for the wave amplitude and a dispersion relationship which allows for the definition of a phase speed (e.g. \cite{thorne}). However, such an analysis does not necessarily extend to the nonlinear theory, as nonlinearities can introduce modified dispersion relations or dissipation mechanisms (compare the Korteweg-de Vries equation \cite{lax}, for example). The analyses of Einstein, Rosen, and others demonstrated that the nonlinearities of the field equations of GR do not allow for phase speeds different from the speed of light \cite{einros,stachel}. Perturbation theory in $f(R)$ gravity, however, demonstrates that gravitational waves have frequency-dependent phase speeds in general \cite{starobinsky} (this is true even in GR with nonzero cosmological constant \cite{robinson,jetzer,bernabeu}). A nonlinear analysis is lacking for the general $f(R)$ theory mainly because of the absence of exact solutions describing gravitational waves \cite{sharyou}. By using the Ernst formalism presented in Section III for cylindrically symmetric, time-dependent metrics, we can construct gravitational wave solutions to the nonlinear theory. Specifically, we construct a solution which has an arbitrary phase speed for a particular choice of $f$. While this does not represent a full treatment of the large-amplitude problem, it does suggest that phase speeds other than the speed of light are possible in $f(R)$ gravity, as the linear perturbation theory in $f(R)$ gravity implies.

Some immediate observations can be made by swapping the variable $\f$ for $\psi$ defined through the relation $\psi = -\tfrac{1}{2}\ln\f$ and letting $B = \rho b$ for some function $b(t,\rho)$. The real and imaginary parts of equation \eqref{eq:timeernst} read, respectively [$f'(R) \neq 0$],
\begin{equation} \label{eq:wave1}
\begin{aligned}
0 =& \left(\sbsn^2 + \frac {\partial_{\rho}} {\rho} \right) \psi - \frac {e^{4 \psi}} {2 \rho^2 b^2} \sbsn \omega \cdot \sbsn \omega - \frac {\sbsn^2 b} {2 b} + \frac {\sbsn b \cdot \sbsn \psi} {b} \\
&- \frac {b_{,\rho}} {\rho b} + \frac {f''(R)} {f'(R)} \left[ \sbsn R \cdot \sbsn \psi - \frac {\sbsn b \cdot \sbsn R} {2 b} - \frac {R_{,\rho}} {2 \rho} \right] ,
\end{aligned}
\end{equation}
and
\begin{equation} \label{eq:wave2}
\begin{aligned}
0 =& \left( \sbsn^2 + \frac {\partial_{\rho}} {\rho} \right) \omega - \frac{ 2 \omega_{,\rho}} {\rho} + 4 \sbsn \omega \cdot \sbsn \psi \\
&- \frac {\sbsn b \cdot \sbsn \omega} {b} + \frac {f''(R)} {f'(R)} \sbsn R \cdot \sbsn \omega ,
\end{aligned}
\end{equation}
which form a coupled set of non-linear hyperbolic wave equations. The second-order piece, $\sbsn^2 + \rho^{-1} \partial_{\rho} = -\partial_{tt} + \partial_{\rho\rho} + \rho^{-1} \partial_{\rho}$, corresponds to the flat-space wave operator in cylindrical coordinates. The fact that the metric functions obey wave equations demonstrates explicitly that $f(R)$ theories predict the existence of gravitational waves \cite{christo,berrygair}. In particular, for the GR case $f(R) = R$, restoring dimensional factors of $c$ shows that the waves propagate at the speed of light \cite{einros}.

Let us now introduce the retarded time $u = t - \kappa \rho$ for some $\kappa >0$, and assume that all metric functions $\psi, \gamma, b$, and $\omega$ are functions of $u$ only, as for a traditional `soliton' solution. The constant $\kappa$ is effectively the phase speed of the gravitational wave; it describes the rate at which disturbances propagate in the spacetime. We confine the metric to the interior of the \textit{causal cone} $C$ given by $C = \{(t,\rho,\phi,z): t \leq \kappa \rho\}$, as is typical of gravitational wave solutions in GR \cite{belzak,tomi,piran,schmidt}. Outside of the causal cone, i.e. for $t > \kappa \rho$, we set\footnote{In general, matching conditions at the boundary of the causal cone impose boundary conditions on the metric functions \cite{israel}. We do not consider the details of the matching procedure here, as they are not germane to the question of the phase speed (however see \cite{gowdy}).} $\gamma = \omega = \psi = 0$ and $b = 1$ (Minkowski space). In this way we construct a spacetime that has a discontinuous wave front representing a propagating gravitational wave in an otherwise empty universe. The metric functions may suffer discontinuities in their derivatives on the boundary of the causal cone like gravitational shock waves (see below). Setting $\kappa$ to unity results in the causal cone coinciding with the light cone. It has been proved that one must have $\kappa = 1$ in GR (e.g. \cite{hawkel}). However non-GR theories may permit $\kappa$ to be either greater than unity (superluminal) or less than unity (subluminal).

Simple solutions of the above form can be constructed by taking 
\begin{equation} \label{eq:powerlaw4545}
f(R) = f_{0} R^{\alpha}.
\end{equation}
For $\alpha > 1$, we find that the Ernst equation \eqref{eq:timeernst}, the Wick-rotated equations \eqref{eq:field5} and \eqref{eq:bfield2} for $\gamma$ and $B$, and the constraint equation \eqref{eq:explicitric} are satisfied for 
\begin{equation} \label{eq:gwomegasol}
\omega(u) = 0,
\end{equation}
and
\begin{equation}
b(u) = \exp\left[ \psi(u) /2 \right],
\end{equation}
provided $\psi$ satisfies the Riccati equation
\begin{equation} \label{eq:psieqn}
0 = 2 \ddot{\psi}(u) - 3 \dot{\psi}(u)^2 - 4 \ddot{\gamma}(u),
\end{equation}
where an overhead dot refers to differentiation with respect to the retarded time $u$. For example, the solution $\psi = \gamma = 0$ yields the Minkowski metric everywhere. It can be easily verified by direct substitution that metrics \eqref{eq:timelineel} satisfying the equations \eqref{eq:gwomegasol}--\eqref{eq:psieqn} solve the $f(R)$ field equations \eqref{eq:fofr1} for any $\alpha > 1$.

If we set 
\begin{equation}
e^{2 \gamma - 2 \psi} = \left( - u \right)^{-2 \beta},
\end{equation}
for some $\beta \geq 0$, equations \eqref{eq:gwomegasol}--\eqref{eq:psieqn} yield the solution
\begin{equation} \label{eq:gwmetrics}
\begin{aligned}
ds^2 =& \left( - u \right)^{-2 \beta} \left[ -dt^2 + d \rho^2 \right] + A_{0} \left( -u \right)^{-1/3 - 2 \sqrt{1-12 \beta}/3}  \\
&\times \left[ (-u) dz^2 + \left( -u \right)^{\sqrt{1-12\beta}} \rho^2  d \phi^2 \right],
\end{aligned}
\end{equation}
where $A_{0} > 0$ is an arbitrary amplitude, which could be fixed by specifying a wave amplitude at some point in space at $t=0$. We have $u < 0$ inside the causal cone, and so the metric \eqref{eq:gwmetrics} is real with Lorentzian signature for $t < \kappa \rho$ provided $0 \leq \beta \leq 1/12$. The metric \eqref{eq:gwmetrics} is singular across the causal boundary $\partial C$ (i.e. $u=0$), as can be seen from the divergence of the $tt$-component of the metric, but it is smooth for all $t < \kappa \rho$. The solution \eqref{eq:gwmetrics} is similar to the simplest Belinski-Zakharov one-soliton solution of GR \cite{belzak,tomi}, which represents the late time behaviour of a particular Einstein-Rosen pulse profile \cite{schmidt,einros}.

While only a toy model which is unlikely to describe a real gravitational wave, the metric \eqref{eq:gwmetrics} demonstrates that the phase speed, $\kappa$, of gravitational waves,  may take arbitrary values in particular $f(R)$ theories. More complicated solutions can be built by considering different functional forms for $f$ using the machinery developed in Sec II and Sec III.


\section{Discussion}
In this paper we derive two generalised Ernst equations for the $f(R)$ theory of gravity in the special cases of stationary and cylindrically symmetric spacetimes. We explicitly derive a class of simple solutions for each case individually and verify that the associated metrics do indeed solve the $f(R)$ field equations. As a physical application, we show that is possible to generalise the Zipoy-Voorhees metric of GR to $f(R)$ theories \cite{zipoy}. The Zipoy-Voorhees metric describes the gravitational field around an ellipsoidal compact body. The generalisation describes a similar object but with some added `hair', i.e. some additional parameters other than mass and angular momentum which appear in the metric coefficients. Additionally, we construct a simple time-dependent metric which seeks to approximate a large-amplitude gravitational wave with arbitrary propagation speed. In GR, it is well known that gravitational waves must travel at the speed of light. However, in an $f(R)$ theory, small-amplitude wave solutions exist which have either sub- or super-luminal propagating wave fronts \cite{starobinsky,robinson,alves}. The small-amplitude result is generalised to arbitrary amplitude here for a particular, time-dependent, cylindrically symmetric metric. Although the result is restricted to this particular metric, it may open a path to more general results in future work.

The $f(R)$ Ernst equations \eqref{eq:ernst} and \eqref{eq:timeernst} offer a few advantages over the usual tensor system \eqref{eq:fofr1}. The Ernst equations, while still nonlinear, are more decoupled than \eqref{eq:fofr1}. The decoupling arises naturally because of the configuration variable approach, which isolates the equations of motion for each metric coefficient. Furthermore, because of the decoupling, there is a sequential recipe for solving these equations, namely for the variable $B$, followed by $\f$ and $\omega$, and finally for $\gamma$. Aside from the practical value in obtaining exact solutions, the Ernst formulation reveals something about the underlying structure of the $f(R)$ field equations. For example, there exists a complex Wick rotation that transforms neatly between solutions for compact bodies and gravitational waves. The Lagrangians associated with the Papapetrou \eqref{eq:lineel} and the Jordan-Ehlers-Kompaneets line elements \eqref{eq:timelineel} are also related by a Wick rotation \cite{beck}, despite having been introduced in different contexts.

We speculate without proof that the formulation presented here extends to the $f(R)$-Maxwell theory, along the lines of Ernst's work on the Einstein-Maxwell theory \cite{ernst2}. If such an extension can be found, it will be interesting to see how the additional nonlinearities in the $f(R)$ field equations interact with the electromagnetic field. Following the outline presented in Section V, it may also be interesting to investigate the properties of gravitational waves in the presence of electromagnetic fields, e.g. in the vicinity of highly magnetised compact objects \cite{magnetosphere,MSM15,MSM17,drumhath}.


Finally, it is worth noting that the Ernst formulation outlined here is \textit{not} unique to the $f(R)$ theory of gravity. Indeed, it applies to any metric theory of gravity that generalises GR and admits a point-like description, for which the procedures outlined in Sections II and III can be replicated. In particular, it can be verified by direct calculation that theories of gravity whose Lagrangian is a function of the curvature invariants $R_{\mu \nu} R^{\mu \nu}$ or $R_{\mu \nu \alpha \beta} R^{\mu \nu \alpha \beta}$, have point-like counterparts independent of $\omega$, i.e. $ \partial \mathscr{L} / \partial \omega = 0$ for either of the parametrisations \eqref{eq:lineel} or \eqref{eq:timelineel}. Such theories include generalised Gauss-Bonnet gravity or the one-loop quantum corrected version of GR \cite{gauss,weyl}.

\section*{Acknowledgements}
We thank Peter Farrell for discussions. We thank the anonymous referee for their helpful suggestions which improved the clarity of this manuscript. This work was supported in part by an Australian Postgraduate Award.


\appendix

\section{Calculation of the Lagrangian $\mathscr{L}$}
The derivation of the Lagrangian \eqref{eq:lagrangian}, for the Papapetrou metric \eqref{eq:lineel}, comes through several applications of integrations by parts, after which total divergence terms are discarded. Noting that $\sqrt{-g} = \f^{-1} e^{2 \gamma} B$ and $\lambda = f'(R)$, the definitions \eqref{eq:action2} and \eqref{eq:explicitric} give us
\begin{widetext}
\begin{equation} \label{eq:aaaaction2}
\begin{aligned}
\mathscr{A} =& \int d^{4} x  \Bigg\{ \frac {e^{2 \gamma} B} {\f} f(R) - \frac {e^{2 \gamma} B } {\f} R f'(R) \\
&- f'(R) \left[  \frac {1} {\f} \bsn B \cdot \bsn \f + \frac {\f^2} {2B} \bsn \omega \cdot \bsn \omega - \frac {3 B} {2 \f^2} \bsn \f \cdot \bsn \f + \frac {B} {\f} \bsn^2 \f - 2 B \bsn^2 \gamma  - 2 \bsn^2 B  \right] \Bigg\}.
\end{aligned}
\end{equation}
\end{widetext}
The action \eqref{eq:aaaaction2} contains second order derivative terms, which must be removed to avoid the Ostrogradsky instability \cite{felice}. In general, we have the elementary formula for well behaved $X$ and $Y$,
\begin{widetext}
\begin{align} \label{eq:someintegral}
\int d z d\rho Y \bsn^2 X &= \int dz d \rho Y  X_{,zz} + \int dz d \rho Y X_{,\rho\rho}  \\ 
&=  \int d \rho Y X_{,z} - \int dz d \rho Y_{,z} X_{,z} + \int dz Y X_{,\rho}  - \int dz d\rho Y_{,\rho} X_{,\rho} \\
&= \int dz d\rho \left[ \left( Y X_{,z} \right)_{,z} + \left( Y X_{,\rho} \right)_{,\rho} - Y_{,z} X_{,z} - Y_{,\rho} X_{,\rho} \right]  . \label{eq:a4}
\end{align}
\end{widetext}
The first two terms in the integrand in equation \eqref{eq:a4} are total divergence terms. Hence, for any $X$ and $Y$, these terms can be removed from the action \eqref{eq:aaaaction2} without modifying the equations of motion \cite{cap1}, i.e. the equations of motion for the Lagrangian
\begin{equation}
\mathfrak{L} =  Y \bsn^2 X,
\end{equation}
are equivalent to those for the Lagrangian
\begin{equation}
\tilde{\mathfrak{L}} = - \bsn Y \cdot \bsn X.
\end{equation}
Making use of relation \eqref{eq:a4} and expanding the integrand in \eqref{eq:aaaaction2} we have

\begin{widetext}
\begin{align}
\mathscr{L} =& \frac {e^{2\gamma} B} {\f} f(R) -  \frac {e^{2 \gamma} B} {\f} R f'(R) - f'(R) \left[ \frac {\bsn B \cdot \bsn \f} {\f}  + \frac {\f^2} {2 B} \bsn \omega \cdot \bsn \omega - \frac {3 B} {2 \f^2} \bsn \f \cdot \bsn \f \right]          \\
&+ \partial_{z} \left[ \frac {f'(R) B} {\f} \right] \f_{,z} + \partial_{\rho} \left[ \frac {f'(R) B} {\f} \right] \f_{,\rho} - 2 \partial_{z} \left[ f'(R) B \right] \gamma_{,z} - 2 \partial_{\rho} \left[ f'(R) B \right] \gamma_{,\rho} - 2  \partial_{\rho} \left[ f'(R) \right] B_{,\rho} - 2 \partial_{z} \left[ f'(R) \right] B_{,z} \\
=& \frac {e^{2 \gamma} B} {\f} \left[ f(R) - R f'(R) \right] + \frac {f'(R)} {2 B \f^2} \left[ 4 B \f^2 \bsn B \cdot \bsn \gamma + \f^4 \bsn \omega \cdot \bsn \omega - B^2 \bsn \f \cdot \bsn \f \right] \\
&+ \frac {f''(R)} {\f} \Big\{ 2 \f \left[ \bsn R \cdot \bsn B + B \bsn R \cdot \bsn \gamma \right] - B \bsn \f \cdot \bsn B \Big\}. \label{eq:somebs}
\end{align}
\end{widetext}
Equation \eqref{eq:somebs} is precisely the form of the Lagrangian \eqref{eq:lagrangian}. Following the procedure presented here, a similar Lagrangian could be derived for the case when the metric variables depend on an arbitrary number of coordinates.

\section{Coordinate independence of the Ernst equations}
Our notation in this article for the operator $\bsn$ differs from Ernst's original presentation \cite{ernst} because we allow for a slightly more general line element in \eqref{eq:lineel} (i.e. we do not demand $B = \rho$). When $B = \rho$, the GR Ernst equation \eqref{eq:grernst} appears to have a coordinate dependency due to the $\rho^{-1} \partial_{\rho}$ term. As Ernst showed, such terms may be removed by introducing the cylindrical 3-gradient $\bsn_{3}$ (as opposed to the 2-gradient $\bsn$) and a new variable $\varphi$ in place of $\omega$ such that terms may be removed to write an equation which respects covariance. For completeness, we show that the same is true for $f(R)$ gravity.

One can re-write equation \eqref{eq:imernst} in terms of cylindrical coordinates $\{\rho,\phi,z\}$ as
\begin{equation}
0 = \bsn_{3} \cdot \left[ \frac {\f^2} {\rho B} f'(R) \bsn_{3} \omega  \right],
\end{equation}
where $\bsn_{3}$ is the usual cylindrical 3-gradient (e.g. \cite{gripod,chandbh}). The well-known identity for any differentiable function $\varphi$ independent of $\phi$ (e.g.  \cite{ernst})
\begin{equation}
0 = \bsn_{3} \cdot \left( \rho^{-1} \hat{\boldsymbol{\phi}} \times \bsn_{3} \varphi \right) ,
\end{equation}
implies that there exists a `potential' $\varphi$ such that
\begin{equation} \label{eq:phirelation}
\frac {\f^2} {B} f'(R) \bsn_{3} \omega = \hat{\boldsymbol{\phi}} \times \bsn_{3} \varphi ,
\end{equation}
where $\hat{\boldsymbol{\phi}}$ is the unit vector in the azimuthal direction. In particular, the relation \eqref{eq:phirelation} is equivalent to
\begin{equation} \label{eq:varphieq}
 \frac {B} {f'(R) \f^2} \bsn_{3} \varphi = -  \hat{\boldsymbol{\phi}} \times \bsn_{3} \omega ,
\end{equation}
which implies that equation \eqref{eq:imernst} may be written as
\begin{equation} \label{eq:imernst2}
0 = \frac {\partial} {\partial \rho} \left[ \frac {B} {\rho f'(R) \f^2} \varphi_{,\rho} \right] + \frac {\partial} {\partial z}  \left[ \frac {B} {\rho f'(R) \f^2} \varphi_{,z} \right] .
\end{equation}
The variable $\varphi$ generalises the quantity introduced by Ernst (see equation (6) of \cite{ernst}) to $f(R)$ gravity. If $B = \rho$, as it must be in GR, equation \eqref{eq:imernst2} reads
\begin{equation}
0 = \bsn \cdot \left[ \frac {\bsn \varphi} {f'(R) U^2} \right],
\end{equation}
which does not contain any coordinate dependent terms. However, in general, $B$ is a function of $\rho$ and $z$ which is unknown \textit{a priori}, so introducing $\varphi$ is unnecessary.

In cylindrical symmetry, the above analysis carries over. We define a potential $\hat{\varphi}$ obeying
\begin{equation} \label{eq:timephi}
 \frac {B} {f'(R) \f^2} \sbsn_{3} \hat{\varphi} = -  \hat{\boldsymbol{\phi}} \times \sbsn_{3} \omega .
\end{equation}
Hence equation \eqref{eq:timefield3}, viz.
\begin{equation}
0 = \sbsn_{3} \cdot \left[ \frac {\f^2} {\rho B} f'(R) \sbsn_{3} \omega  \right],
\end{equation}
is equivalent to
\begin{align} \label{eq:imtimeernst2}
0 &= \frac {\partial} {\partial \rho} \left[ \frac {B} {\rho f'(R) \f^2} \hat{\varphi}_{,\rho} \right] - \frac {\partial} {\partial t}  \left[ \frac {B} {\rho f'(R) \f^2} \hat{\varphi}_{,t} \right] \\
&= \sbsn \cdot \left[ \frac {\sbsn \hat{\varphi}} {f'(R) \f^2} \right].
\end{align}
If one is interested in $f(R)$ solutions such that $B = \rho$ is fixed, substituting the variable $\varphi$ through \eqref{eq:varphieq} or its Wick-rotated counterpart $\hat{\varphi}$ through \eqref{eq:timephi} ensures that the resulting equations are coordinate insensitive.

\end{document}